# Towards grid-enabled telemedicine in Africa

Florence JACQ[1], Frank BACIN[2], Nonfounikoun MEDA[3], Denise DONNARIEIX[4], Jean SALZEMANN[1], Vincent VAYSSIERE[1], Nicolas JACQ[1], Michel RENAUD[5], François TRAORE[6], Gertrude MEDA[6], Rigobert NIKIEMA[6], Vincent BRETON[1]

[1]*Laboratoire de Physique Corpusculaire (CNRS/IN2P3), Campus des Cézeaux, Aubière, 63177, France*
Tel: +33473407849, Fax: +3373264598, Email: fjacq@clermont.in2p3.fr
[2]*CHU Gabriel Montpied, 58 rue Montalembert, Clermont-Ferrand, 63000, France*
Tel: +33473750750, Email: fbacin@chu-clermontferrand.fr
[3]*CHU Yalgado Ouedraogo, Ouagadouou, Burkina Faso*
Email: nd.meda@univ-ouaga.bf
[4]*Centre Jean Perrin, Clermont-Ferrand, 63000, France*
Tel : +33473278375, Fax+33473278125
Email: denise.donnarieix@cjp.fr
[5]*Cabinet d'ophtalmologie, Cournon, 63800, France*
Email : michel-renaud2@wanadoo.fr
[6]*Centre Médical Schiphra, Ouagadougou, BurkinaFaso*

**Abstract:** Telemedicine services are very relevant tools to train local physicians and to improve diagnosis by exchanging medical data. Telemedicine networks allow these exchanges but the set-up of multipoint dynamic telemedicine requires moving towards GRID technologies. A healthgrid is an environment where data of medical interest can be stored and is easily available between the different actors of healthcare. Two telemedicine applications were developed to link physicians from Burkina Faso and France with the perspective of setting up a grid infrastructure between the participating medical sites. A web site to exchange diagnosis on diabetic retinopathy was developed in PHP and another application using web services was developed to exchange patient information between two databases.

**Keywords:** Telemedicine – GRID – Web services – Diabetic Retinopathy– Ophthalmology – Burkina Faso – Exchange of diagnosis.

## 1. Introduction

*1.1 The challenges of telemedicine for medical development*

The concept of telemedicine represents all the potential means to exchange medical data or images whatever the distance between two physicians. The sharing of knowledge and the exchange of diagnosis between physicians is the key to improving the standard of medical knowledge, particularly in developing countries [1]. Today, information technologies allow efficient virtual links between two groups of healthcare professionals in the world.

For historical reasons, physicians of several African countries have established strong relationships with French colleagues for many years. Telemedicine allows the pursuit of this collaboration effectively by strengthening the link between French clinicians and African physicians when they return to their home country. Telemedicine provides a way to collect and exchange medical data, to share diagnosis and therefore to improve the training of physicians.

As stressed in a very recent report from the European Commission and the World Health Organization, "Information and communication technologies are changing health care delivery and are at the core of effective, responsive health systems. These technologies are key to connecting people, information and research to improve health in countries. They are also vital in enabling rapid response to global threats to health." [2]. Efficient eHealth services have already demonstrated their value [3]. But the challenge is to widen their implementation and adoption particularly in developing countries.

Physicians can take advantage of development of Internet for telemedicine services [4-7]. For instance, teleopthalmology services seem to be adapted for diagnosis and patient follow-up [8]. An effective architecture was also deployed for remote diabetic retinopathy screening [9].

But with this technology, the user has to look after the information, to upload images on a server or to download patient information from a server. Thus, users have to take charge of all the different steps of information exchange.

*1.2 Telemedicine networks*

Most of the telemedicine projects are designed to allow the exchange of information between two groups of healthcare professionals, in developed and/or developing countries. Such an exchange is very well fitted to provide a second diagnosis. However, the set-up of multipoint dynamic telemedicine networks where several teams could share patient data while respecting patient privacy requires the further strengthening of collaboration between healthcare professionals even within developing countries. In these networks, the medical teams in healthcare centres would be capable of sharing medical data on demand for second diagnosis [10]. This requires moving away from the present central web server approach towards the creation of a federation of databases where data is stored in the healthcare centres and made available for punctual enquiries whilst safeguarding patient privacy.

Such networks would be efficient tools to collect epidemiological data in relation to neglected diseases. Indeed, epidemiologists need to collect data from endemic areas to develop models of a given disease, to evaluate drug resistance, to monitor drug distribution, etc. The medical data available in hospitals geographically distributed in endemic areas is extremely relevant provided it can be consulted for epidemiological analysis. GRID technology allows the creation of such a federation of databases.

*1.3 The GRID impact in telemedicine*

The application of grid technology to healthcare has been explored within the framework of a white paper introducing and explaining the concept of healthgrid [11]. A healthgrid is an innovative use of information technology to support broad access to rapid, cost-effective and high quality healthcare. An emerging aspect of eHealth, it is an environment where data of medical interest can be stored, processed and made easily available to the different actors of healthcare, physicians, healthcare centres and administrations, and of course citizens.

The recent emergence of grid technology opens new perspectives to enable telemedicine and medical research with developing countries, particularly for preparation and follow-up of medical missions as well as support to local medical centres in terms of teleconsulting, telediagnosis and patient follow-up. Indeed, grids mask the complexity of handling distributed data in such a way that physicians will be able to access patient data without being aware where this data is stored.

GRID technology opens perspectives to empower existing telemedicine services in terms of interactivity and access to data. It allows the exchange of patient data without moving it on Internet networks. Patient information becomes accessible from any GRID site provided that access rights are given.

*1.4 Towards grid-enabled telemedicine for Africa*

In the last 5 years, our research group in Clermont-Ferrand (http://clrpcsv.in2p3.fr) has been developing life sciences and healthcare applications on the grid infrastructures deployed within the framework of the DataGrid (FP5) and EGEE (FP6) European projects. Our objective is to develop grid-enabled telemedicine services for medical development in collaboration with Non Profit Organizations in China and in Burkina-Faso.

Our approach is first to offer conventional internet-based telemedicine services as a first step towards a grid-enabled service while making sure the architecture and design of the services make it easy to deploy in a grid environment. We are developing our own telemedicine environment with web services to answer specific requirements of physicians involved in our collaborations. All development and deployment choices are made with the perspective to migrate on a grid solution once physicians have adopted the telemedicine usage.

In this article we present the current deployment in Burkina Faso of two telemedicine applications based on the Internet with the idea to connect them in the future with the GRID. Objectives are described in the first part. The methodology is then explained in the second part. The third part focuses on the description of the technology. The fourth part concerns the applications development. The work finishes with the presentation of the results .

## 2. Objectives

We started in China where telemedicine services were set-up to support a medical development project of the French NPO Chain of Hope [2]. A first protocol was established for describing the patients' pathologies and their pre- and post-surgery states through a web interface in a language-independent way. This protocol was evaluated by French and Chinese clinicians during medical missions in the fall of 2003.

We have very similar aims in Burkina-Faso where the French NPO Les Eaux Vives has been supporting for 2 years the development of an ophthalmologic surgery unit at the Schiphra dispensary. Teams of French clinicians go regularly to Burkina-Faso to perform surgery and to train local healthcare professionals. In parallel, collaboration has been established for several years between Clermont-Ferrand and Ouagadougou hospitals. The teams of clinicians involved expressed a strong interest in telemedicine services to facilitate information exchange between France and Burkina-Faso.

Specific requirements concerning interfaces or the software used led us to develop new telemedicine services based on the Internet which could be connected to the grid as a next step. Physicians need tools to train efficiently local physicians and to exchange diagnoses in order to improve the quality of services. These tools have to be easy to use and have to offer a user-friendly interface for physicians. Two telemedicine projects were initiated with Burkina Faso:
- Development of an application for training and diagnosis exchange on diabetic retinopathy between Clermont-Ferrand (France) and Ouagadougou (Burkina Faso) hospitals
- Deployment of software to improve medical support to a dispensary in Ouagadougou

*2.1 Telemedicine website for diabetic retinopathy*

Diabetic retinopathy is a disease induced by diabetes which is more and more frequent in Africa. The difficulty lays in the disease diagnosis and the choice of the best treatment. Two clinicians, an ophthalmologist based in Clermont-Ferrand and an ophthalmologist based in Ouagadougou have expressed a need to build an application to allow the French physician to give a second diagnosis for patients in Burkina Faso. Some forms exist in the literature to guide the physician in his diagnosis and to find the best treatment for diabetic retinopathy. The aim of this application was to computerize these forms and to store patient data and images. The application developed could be potentially support local training.

*2.2 Telemedicine application for the management of medical data*

"Les Eaux Vives", a non-profit humanitarian French association created in 1978, has been providing support to the development of ophthalmologic surgery at Schiphra dispensary in Ouagadougou since 2004. The association coordinates regular missions with European healthcare professionals to assist local clinicians who operate cataracts. As these missions are limited in time, it was perceived that a computerized infrastructure would help the management of the patients and the exchange of information between French and local clinicians. StudioVision (http://www.realvision.fr/html/oph.htm), commercialized by RealVision, is software which takes charge of all the tasks surrounding the medical act: agenda, bookkeeping, prescription etc. Moreover, with the installation of Studio Vision in the dispensary, it is possible to exchange information between the Studio Vision database in Ouagadougou and Clermont-Ferrand. In other words, what an ophthalmologist registers using Studio Vision in Ouagadougou is available in the StudioVision database in Clermont-Ferrand. It is a way to exchange diagnosis between French ophthalmologists and the dispensary.

## 3. Methodology

As we develop informatics solutions for a distant country, the applications have to be portable and secure in order to respect patient privacy in the exchanges. They must also take into account the working habits of each physician.

For the diabetic retinopathy, physicians use a sheet of paper to fill in the standard forms to diagnose the diabetic retinopathy. So, the solution proposed for this disease was to create a website where the standard forms are accessible securely by ophthalmologists. The website can store the medical data and allow the exchange of diagnosis.

To improve patient data management in the dispensary, the solution was to deploy Studio Vision on machines and to develop an application which communicates with Studio Vision in order to transfer data from Burkina-Faso to France. The information flow between two distant Studio Vision databases is presented in figure 1.

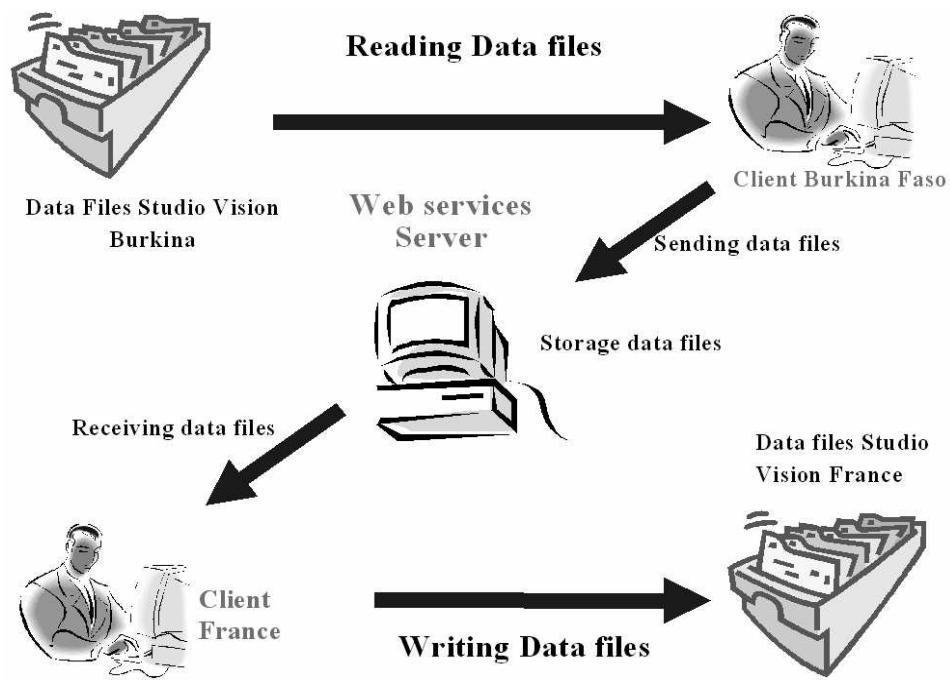

*Figure 1: Information flow between two distant Studio Vision databases*

In these two cases, a model was built in France which was tested by physicians before a deployment in Burkina Faso was scheduled. The deployment of these applications is followed by a medical training based on these applications.

## 4. Technology Description

The telemedicine services were developed with the idea to connect them in the future with the GRID. Therefore they were conceived to be totally compatible with a GRID infrastructure. Technologies used for these applications are free, standard and easily interoperable with the GRID. The two applications are secured with a login and password.

The web site was developed in PHP and connected to a MySQL database to store the medical data. In order to respect the French laws regarding protection of patient privacy, the medical data is encrypted with an encrypting algorithm.

The application connected to Studio Vision was developed in Java, so that it can work with all different operating systems. Grid middlewares are progressively moving from classic network programming to a more flexible standard brought by the web services (like Globus Toolkit 3, Globus Toolkit 4 [12], Rugbi (http://rugbi.in2p3.fr/), Glite (http://glite.web.cern.ch/glite)). We developed internal services of the applications using web services technologies communicating with each other via SOAP protocol. SOAP uses XML messages encapsulated in http packets, so all the messages contents are also in XML. The patient data stored in the Studio Vision database (Access Database) are extracted through SQL queries and formatted into XML files. Then these structured metadata can be easily transferred by SOAP queries which are well adapted for this purpose, as this data is of relatively small size, basically a few tens of kilobytes. Moreover the chosen architecture enables any client to use the application, whatever his network configuration is. Web services containers are running on http servers, hence a server only needs to open up inbound connectivity on the http or https ports. A client on the other hand can reach the web services containers and communicate with them using SOAP with an internet connection, even behind routers or firewall. All the clients can work with each other asynchronously reaching for the server to send or retrieve data. This is exactly the kind of functioning we can observe in a grid environment: users calling remote services providing storage and computing resources. The developed web services could be easily deployed under Grid-services containers without any change of the server software environments.

## 5. Developments

The diabetic retinopathy web site allows users to register medical data and attach images for a single patient. This data is anonymous and stored in a database. The medical data is registered according to the standard form used by ophthalmologists.

To deploy the software Studio Vision in the Schiphra dispensary, it was necessary to have at least three computers, one computer for the Studio Vision server and two used by two doctors. A telemedicine application was developed to transfer medical data registered in Studio Vision in the Ouagadougou dispensary to the Studio Vision database in Clermont-Ferrand. The application is split in two parts, a "send" part for the dispensary and a "receive" part for the French physician.

Three computers are ready to be sent and installed in Ouagadougou. They will be shipped together with medical equipment for the ophthalmologic surgery unit in the spring 2006 by the "Les Eaux Vives" association.

## 6. Results

The web site was presented in the Hospital Yalgado OUEDRAOGO of Ouagadougou to the ophthalmologists in December 2004. The web site was tested using data of a real patient. This first test allowed us to improve the forms to ease the task of registration by the physicians. The only difficulty was to upload images as the network was limited: low bandwidth, high latency and high loss. This problem was already encountered in our China experience. But the service can be used despite the slow upload. Unfortunately the Hospital has not enough money to buy the equipment necessary to have some specific pictures of the eye called angiographies. So for the moment, the web site is available but it cannot be used by the physicians of the hospital. Collaboration between physicians has stayed at the same level as before (email exchange). However, we will propose this site to the ophthalmologists of the dispensary before the summer 2006.

The Eaux Vives association donated three computers for installation in the Schiphra dispensary. The StudioVision software was installed as well as the telemedicine application. A first test has been carried out in Clermont-Ferrand by simulating the patient data transfer between two different points. The computers should be shipped in the spring of 2006. Their deployment will also be the opportunity to evaluate network performances and therefore the feasibility of installing a grid node at the Schiphra dispensary. In parallel, we are developing a new generation of telemedicine services using experience acquired in the last two years.

Implementation and adoption problems were analysed to deploy telemedicine services. First of all, people involved in the projects are convinced of the utility of the telemedicine. The telemedicine applications were built to be easily integrated into the dispensary or the hospital especially when it comes to getting patients in for examinations, managing their records and conducting follow-up. Initial technical shortcomings do not dampen the enthusiasm for the project. The main difficulty comes from limited resources in developing countries including staff skills, bandwidth and funding. But the collaboration with dispensaries can open a door to enhanced collaboration between physicians of France and Burkina Faso.

## 7. Conclusions

Telemedicine applications are being developed within the framework of collaboration between ophthalmologists in Burkina Faso and France to foster medical development in Burkina-Faso. One of the aims of exchanging patient information between two medical databases is to facilitate the follow-up in an ophthalmologic surgery unit newly created at the Schiphra dispensary in Ouagadougou. The service will be deployed before the summer of 2006. Thus we will analyse the implementation difficulties. The other is a web site designed for the exchange of diagnosis on diabetic retinopathy between ophthalmologists of Clermont-Ferrand and Ouagadougou hospitals. Due to the lack of equipment, the deployment in the Ouagadougou hospital is suspended for an undetermined time. But there are new perspectives to deploy it in the dispensary. Resources, like equipment and network, do not limit the implementation.

These applications are designed with the aim of their use on grids which opens the perspective of multipoint dynamic telemedicine. We are developing a new generation of telemedicine service using experience acquired in the last two years.